\documentclass[review,preprint]{revtex4}
\usepackage[dvips]{graphicx}
\usepackage{t1enc,amsmath}
\usepackage{amsfonts}
\usepackage{amssymb}
\begin{document}

\title{Quantification of the transmission properties of  metasurfaces illuminated by finite-size beams}
\author{M. Boutria$^{1,2}$, A. Ndao$^{1,3}$ and F. I. Baida$^{1,*}$}

\affiliation{$^1$ Institut FEMTO-ST, UMR CNRS 6174, Universit\'e Bourgogne Franche-Comt\'e, 25030 Besan\c con, France\\ $^2$ Ecole Normale Sup\'erieure de Kouba, B.P. 92 - 16050 Alger - Algeria\\
$^3$ Department of Electrical and Computer Engineering, University of California San Diego, La Jolla, California 92093-0407, USA}


\begin{abstract}
The aim of this letter is to  {present analytical method to} quantitatively address the influence of a focusing illumination on the transmission {/reflection} properties of a  {metasurface illuminated by a finite-size beam}. In fact, most theoretical and numerical studies are performed by considering an infinite periodic structure illuminated by a plane wave.  {In practice}, one deals with a finite-size illumination and structure. Combination of the angular spectrum expansion with a monomodal modal method is performed to determine the beam size needed to {acquire} efficient properties of  {a Metasurface that behaves as  Anisotropic Plate (\textbf{MAP})}. Interesting results show that the beam-size can be as small as $5\times5$ periods to recover the results of a plane wave. Other results also show that the beam-size could be used as an extrinsic parameter to enhance the MAP performance and  {to finely adjust} its expected properties (birefringence and/or transmission coefficient).
\end{abstract}

\maketitle

More often than not, theoretical studies dealing with the design of periodic metamaterial exhibiting original properties (extraordinary or enhanced transmission, large anisotropy, enhanced nonlinearity, chirality, {etc}) involve {an} infinite periodic structure illuminated by a plane wave \cite{Martin-Moreno:prl03,Popov:prb00,cao:prl02,baida:oc02}. This approach is questionable when dealing with finite size optical beams and/or finite size structures. This is particularly critical when the properties of the metamaterial are very sensitive to the illumination angle of incidence. This especially  {happens when} surface plasmon {resonances are involved}.  {In the latter case}, it was found that a structure size  {larger} than the propagation length of the surface plasmon is generally needed to recover the properties of an infinite size structure \cite{Przybilla:oe08}. {Guizal et al. \cite{Felbacq:08}}, studied the propagation of a finite size beam through a 2D metamaterial (slabs) {.  They show that the group velocity direction can be different from the normal to the iso-frequency curves of a dispersion diagram} due to an efficient contribution of the evanescent waves that are excited inside and at the edges of the beam. Unfortunately, there are only a few studies where  {extensive} numerical simulations are performed using commercial softwares by considering  {finite-size} structure and/or  {beams }\cite{boye:ao00,lin:nl10,giovampaola:natmat14,kivijarvi:njp15}. These simulations are generally essential so as to model non-periodic metamaterials \cite{lieven:nl09,capasso:nl16}. Fortunately, there are some simpler issues to treat the case of periodic structures. A. Roberts \cite{roberts:oe10} studied the transmission of different polarization state beams through an array of coaxial apertures engraved into a perfectly conducting screen. An angular spectrum expansion was used in that study but the mean beam propagation direction was restricted to normal incidence. Here {,} we propose to combine the angular spectrum expansion with a monomodal modal method to determine the transmission properties of a specific  {metasurface} behaving as half-wave or quarter-wave  {plate} \cite{baida:prb11}. The main objective of this work is to  {show that this method can} provide the optimal experimental conditions (structure and beam dimensions,  {angle of incidence}) for which the structure operates effectively.
\begin{figure}[ht]
\centering
\includegraphics[width=10cm]{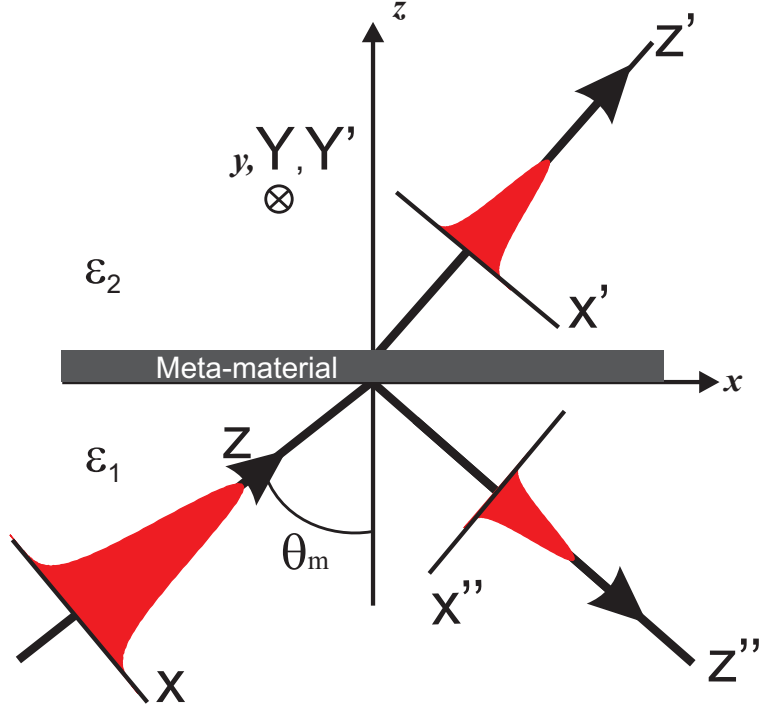}
\caption{Schematic of the model with different frames involved in the theoretical model.}
\label{objet7}
\end{figure}

The incident beam is a fundamental Hermite-Gauss beam ($HG_{00}$) supposed to propagate in the $xOz$ plane along the {$Z$-axis} located at the angle $\theta_m$ from the normal to the  {metasurface}. For the three fundamental polarization states (circular, elliptical or linear), its angular spectrum expansion can be expressed in the $Oxyz$ reference frame related to the  {metasurface} (see fig. 1) by:
\begin{eqnarray}
F_{x}(k_x,k_y) &=& A_{inc}(k_x\cos \theta_m +k_z\sin \theta_m ,k_y)\left\{\alpha-\frac{k_y\sin\theta_m}{k_z}\beta\right\} \notag
\\ F_{y}(k_x,k_y) &=&A_{inc}(k_x\cos
\theta_m +k_z\sin \theta_m ,k_y)\left\{ \beta\left( \cos \theta_m
+\dfrac{k_x\sin \theta_m }{k_z}\right) \right\}  \label{champ}\\
F_{z}(k_x,k_y) &=&-A_{inc}(k_x\cos \theta_m
+k_z\sin \theta_m ,k_y)\left\{ \dfrac{k_x\alpha+\beta k_y\cos \theta_m}{%
k_z}\right\} \notag
\end{eqnarray}
Where $A_{inc}(k_X,k_Y)=\pi W_0^{2}E_{0}\exp \left[-W_0^{2}(k_X^{2}+k_Y^{2})/4\right]$ is the Fourier amplitude of the plane wave characterized by its $(k_X,k_Y)$ transverse wave vector components expressed in the proper frame of the Gaussian beam ($OXYZ$). $k_x$, $k_y$ and $k_z$ are the same components in the frame of the structure, $E_0$ is the maximum amplitude of the electric incident field and $W_0$ is the beam-waist defined as the beam width at $1/e$ of its amplitude $E_0$. $\alpha$ and $\beta$ are given by:
\begin{eqnarray}
\alpha&=&cos\phi \notag
\\ \beta&=&a\sin\phi \notag
\end{eqnarray}
where $\phi$ and $a$ are two parameters that define the polarization state of the incident beam as:
\begin{itemize}
\item linear directed along the angle $\phi$ measured from the $X$-axis with $a=1$ (TE for $\phi=\pi/2$ and TM for $\phi=0$)
\item circular with $\phi=\pi/4$ and $a=\pm\sqrt{-1}$
\item elliptical with major to minor axes ratio equal to $|\beta/\alpha|=tan\phi$ and $a=\pm\sqrt{-1}$ (ellipse axes along the $x$ and $y$ directions).
\end{itemize}

This formulation is limited to an illumination direction in the $xOz$ plane and need a simple variable change in the case of a tilted plane of incidence.

On the other hand, the monomode modal method was  {explained} in details in ref. \cite{baida:prb11} where a numerical code was developed  {to determine} the transmission and reflection Jones matrices in the $(TE,TM)\equiv(s,p)$ basis for an illumination characterized by the angle of incidence $\theta$ and the azimuthal angle $\psi$ through:
\begin{equation}
 t(\theta,\psi)= \left(\begin{array}{cc}
 t_{ss} & t_{sp}
 \\t_{ps} &  t_{pp}
 \end{array}\right)
 \end{equation}
 Transmission amplitude of the whole beam is then the sum of all transmission coefficients (in amplitude) of each plane wave defined by its tangential wave vector $(k_x,k_y)$. To evaluate this sum, we  {first need} to express the Jones matrix of each plane wave as a function of its wave vector by linking the angles $\theta$ and $\psi$ to $k_x$ and $k_y$:
\begin{eqnarray}
\theta &=& \cos^{-1}(\frac{k_z } {n_1k_0})\\
\psi &=& \cos^{-1}(\frac{k_x }{ k_{\parallel}})\notag
\end{eqnarray}
where $k_z=\sqrt{\varepsilon_1 k_0^2-k_{\parallel}^2}$ with $k_{\parallel}$ is the tangential component of the wave vector ($k_{\parallel}=\sqrt{k_x^2+k_{y}^2}$).
Finally, a basis change from $(x,y)\rightarrow (s,p) \rightarrow (x,y)$ is necessary to determine the transmitted electric field ($\overrightarrow{F_t}(k_x,k_y)$) components in the $Oxyz$ frame of the  {metasurface}:

\begin{equation}
\overrightarrow{F_t}(k_x,k_y)=\wp^t\cdot t(\theta,\psi)\cdot\wp \cdot\overrightarrow{F}(k_x,k_y)
\end{equation}
the $^t$ denotes the transpose matrix operator and $\wp$ is the basis change matrix $(s,p)\rightarrow (x,y,z)$  given by:

 \begin{equation}
 \wp(k_x,k_y)= \left(\begin{array}{cc}
-\cos\psi\cos\theta & \sin\psi\\
-\sin\psi\cos\theta & -\cos\psi
\\ \sin\theta & 0
 \end{array}\right)
 \end{equation}
 The transmitted zero order amplitude of the whole beam in a $xOy$ plane located at $z=z_0>>\lambda$ (in the far-field) is then calculated by integrating the transmitted amplitude issued from all the incident plane waves. This leads to the following angular spectrum expansion:
 
 \begin{equation}
 \overrightarrow{E_t}(x,y,z_0)=\cfrac{1}{4\pi^2}\int \hspace{-.3cm}\int \limits_{-\infty}^{+\infty} \overrightarrow{F}_{t}(k_x,k_y)e^{-ik_x x-ik_y y-ik_z z_0}dk_x dk_y
 \end{equation}

We propose here to employ this formalism to point out the transmission properties of  {a MAP}. Consequently, we will define a criterion to evaluate the deviation of the transmitted beam properties from those of an incident plane wave. This is achieved through the determination of three parameters that  {will help} to quantify the transmitted beam properties  {and are, as well,} directly related to experimental considerations involving quantification {obtained} through simple experiments. Let us enumerate these parameters:\\

\textbf{1)} $T$: The transmission coefficient defined as the ratio of the transmitted total power divided by the incident one.

\begin{equation}
  T=\frac{\displaystyle \int \hspace{-.3cm}\int \limits_{-\infty}^{+\infty} \vec{E}_t(x,y,z)\wedge\vec{H}_t(x,y,z)dx dy}  {\displaystyle \int \hspace{-.3cm}\int \limits_{-\infty}^{+\infty} \vec{E}_i(x,y,z)\wedge\vec{H}_i(x,y,z)dxdy}
  \end{equation}
This parameter is obviously very important and must be maximized ($100\%$) in order to build efficient optical components.\\

\textbf{2)}  $\xi$: The average phase change induced between the two transverse components of the transmitted beam. This parameter depends on the awaited  {metasurface} properties. In our case, we are dealing with $\lambda/2$ ($\lambda/4$) plate so we expect a target value of $\xi=\pi$ ($\xi=\pi/2$) between the two transverse components of the electric field directed along the plates axes ($Ox$ and $Oy$). Thus, we define this parameter as:
    
  \begin{equation}
  \xi=\frac{1}{S}\int\limits_{(S)} \hspace{-.2cm} \int Arg \left(\cfrac{{E}_{tX'}(X',Y',Z'=z_0)}{{E}_{tY'}(X',Y',Z'=z_0)}\right)dX'dY'\label{xi}
  \end{equation}

Where the integration is numerically restricted over a surface ($S$) corresponding to the set of points ($X',Y'$) in the plane $Z'=z_0$ such that $I_t(X',Y',z_0)\geq I_{max}10^{-2}$. For an HG$_{00}$ beam, this corresponds to $99.08\%$ of the total energy. The frame $OX'Y'Z'$ is related to the transmitted field and is the same as $OXYZ$ when $\varepsilon_1=\varepsilon_2$. Otherwise, a frame change (rotation induced by the light refraction) is required.
As defined by eq. \ref{xi}, the numerical value of the phase shift is determined between $-\pi$ and $\pi$. Thus {,} we have to be careful when making the sum of all   {values} by assuming a continuous spatial phase change (unwrapping the results before making the summation).\\

\textbf{3)}  The polarization of the transmitted beam can be quantified by determining the spatial average of the polarization degree $P$ calculated as \cite{wolf:nc59}:

\begin{equation}
P=\left<\sqrt{1-\cfrac{4|J|}{(J_{xx}+J_{yy})^2}}\right>_{x,y,z_0}
\end{equation}
Where $J$ is the polarization matrix or the coherency matrix defined by \cite{qasimi:ol07}:

\begin{equation}
J= \left(\begin{array}{cc}
J_{xx} & J_{xy}
 \\J_{yx} & J_{yy} \end{array}\right)=\left(\begin{array}{cc}
<E_xE^*_x>_t & <E_xE^*_y>_t
 \\<E_yE^*_x>_t& <E_yE^*_y>_t
 \end{array}\right)
 \end{equation}

$|J|$ is the matrix $J$ determinant.

The value of $P$ is real positive  {$\in[0;1]$}.  {$P=0$ corresponds to a completely unpolarized beam while the latter is entirely polarized (linear, circular or elliptical) when $P=1$}. Moreover, we are looking for MAP that modifies the polarization direction ($\lambda/2$ plate) or the polarization nature ($\lambda/4$ plate). In both cases, the transmitted beam must be completely polarized whatever the polarization of the incident beam. This means that the value of $P$ must be  {close} to $1$.

According to these parameters ($T, \xi$, and $P$) and  {for anisotropic plates}, we define a {figure-of-merit $FOM$} function that corresponds to the total deviation between the response of a plane wave from that of the Gaussian beam by:
\begin{equation}
 {FOM}=\Delta T+\Delta P+\Delta(\cfrac{\xi}{\Psi})  \label{FOM}
\end{equation}
where $\Psi$ is the phase change expected to be introduced by the MAP, the operator $\Delta$ denotes the deviation with respect to the plane wave case. Consequently, we assume that a maximum value of  {$FOM_{max}=10^{-2}$} is allowed to recover the plane wave behavior. This value can be seen as an onerous condition. Nevertheless, we see in the following that this criterion can be easily fulfilled in the case of lossless materials (perfect conductors for instance).
\begin{figure}[ht]
\centering
\includegraphics[width=12cm]{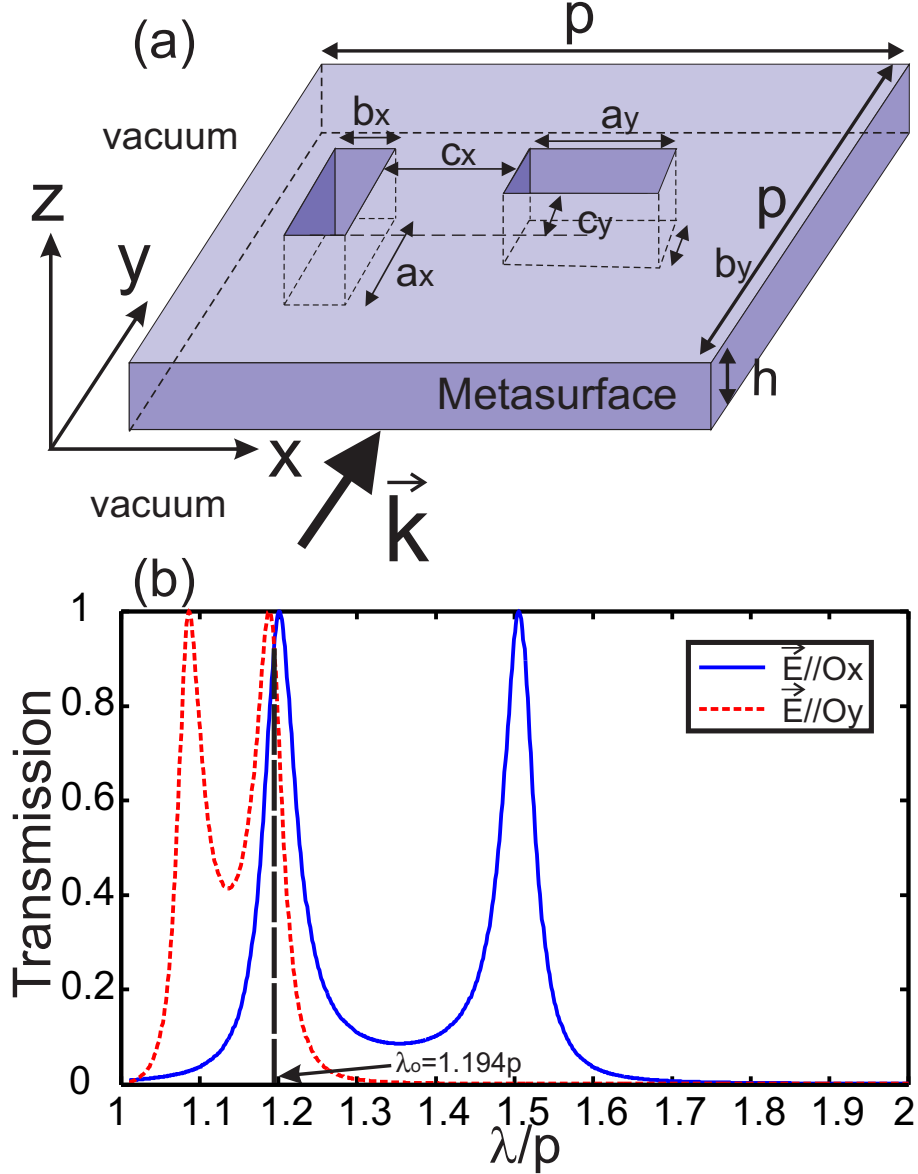}
\caption{(a) Schema of  {a unit cell of the MAP}. The rectangular apertures are engraved in a perfectly electric conductor layer of thickness $h$. The two rectangle dimensions are different in order to induce artificial anisotropy. (b) Typical transmission spectra for the two orthogonal polarization states along $OX$ and $OY$ when {$\theta_m=0^\text{o}$} ($OX\equiv OX' \equiv OX'' \equiv Ox$). }
\label{schemalame}
\end{figure}
Let us now consider the two MAPs studied  {on Figs. 4(a) and 4(b) of ref. \cite{baida:prb11} corresponding to half-wave }and quarter-wave plates respectively. The general schema of such MAP is depicted in Fig. \ref{schemalame} and consists on a bi-periodic grating (period $p$ along both $x-$ and $y-$directions) of two rectangular apertures engraved within an $h$-thick layer of perfect electric conductor. All the geometrical parameters are explicitly given in the caption and in the insets of  {Figs. 4(a) and 4(b)} of ref. \cite{baida:prb11}. Nonetheless, they will be reminded below for each case.

\begin{figure}[ht]
\centering
\includegraphics[width=12cm]{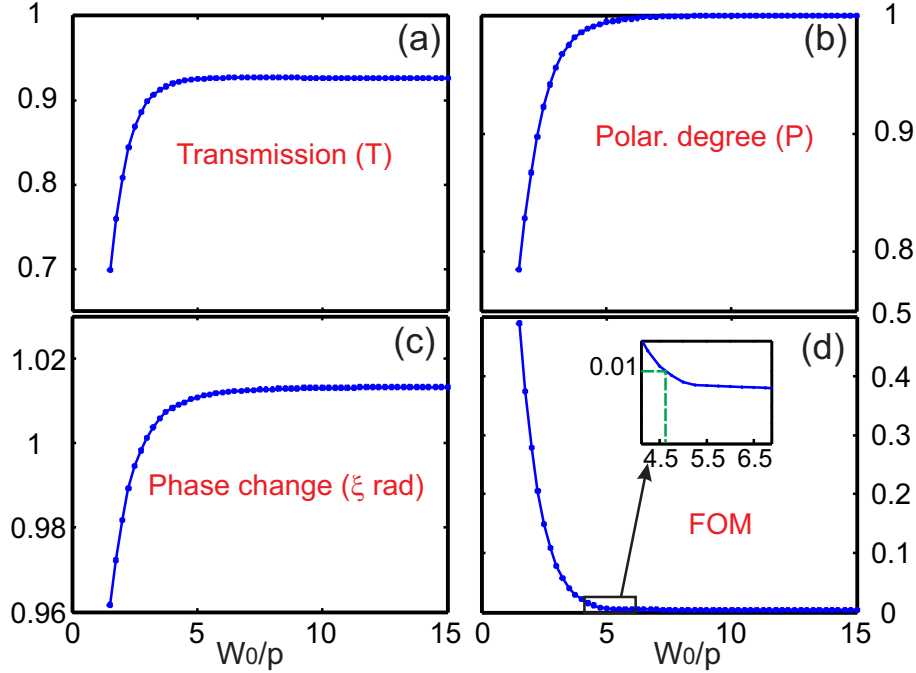}
\caption{Variations of (a) the transmission efficiency $T$, (b) the polarization degree $P$, (c) the phase change $\xi$ and (d) the  {$FOM$} defined by equation \ref{FOM} in the case of the half-wave plate operating at $\lambda=1.194p$ as a function of the illumination Gaussian beam size $W_0/p$. The beam is impinging the structure under normal incidence and is polarized at  {$45^\text{o}$} from the $x-$axis in order to get transmission across the two perpendicular rectangles. The plane wave expansion of the beam is described with  {$256\times256$} harmonics. The inset in {Fig.} (d) corresponds to a zoom made  {near the point $FOM=10^{-2}$.}}
\label{demionde}
\end{figure}

Let us first consider the case of the half-wave  {plate} where $a_x=0.75p, b_x=0.1p, c_x=0.15p, a_y=0.5818p, b_y=0.2p, c_y=0.6p$ and $h=0.8p$. When illuminated by a plane wave at normal incidence, the transmission coefficient, the phase change, and the polarization degree are  {$T_{\text{pw}}=0.9256$, $\xi_{\text{pw}}=1.01\pi$ and $P_{\text{pw}}=1$} respectively at the operation wavelength $ {\lambda_0}=1.194p$. As shown  {in Fig.  \ref{schemalame}(b)}, this wavelength corresponds to intersection between the transmission spectra of a $x-$ and $y-$ polarized incident plane waves.\\
Figure \ref{demionde} presents the variations of the three parameters $T$ (a), $\xi$ (b) and $P$ (c), {and the $FOM$} as a function of the beam-waist $W_0$ when this MAP ($\lambda/2$ plate) is illuminated by a linearly polarized Gaussian beam. Due to the  {structure} anisotropy, these parameters will depend on the polarization direction of the incident beam especially for highly focused beams (small $W_0$). To take into account this anisotropy, we choose to fix polarization of the incident beam to  {$\phi=45^\text{o}$}.  As we can see, all  {of} the three parameters tend asymptotically to  {a value} corresponding to  {that} of a plane wave illumination. A  {$FOM$} of $10^{-2}$ is reached for a beam-waist $W_0^{min}=5p$ (see inset of  {Fig. \ref{demionde}(d)}. The  {latter} corresponds to $P=0.9944$, $\xi=1.011\pi$ and a transmission efficiency of $T=0.9252$. This result demonstrates the robustness of the structure with respect to the beam-size. In fact, one might expect qualitatively this outcome because of the underlying physical effect  {(the phase change)} is not a collective effect but it is induced by every couple of apertures (each period) through the excitation of a guided mode inside the apertures.

\begin{figure}[ht]
\centering
\includegraphics[width=12cm]{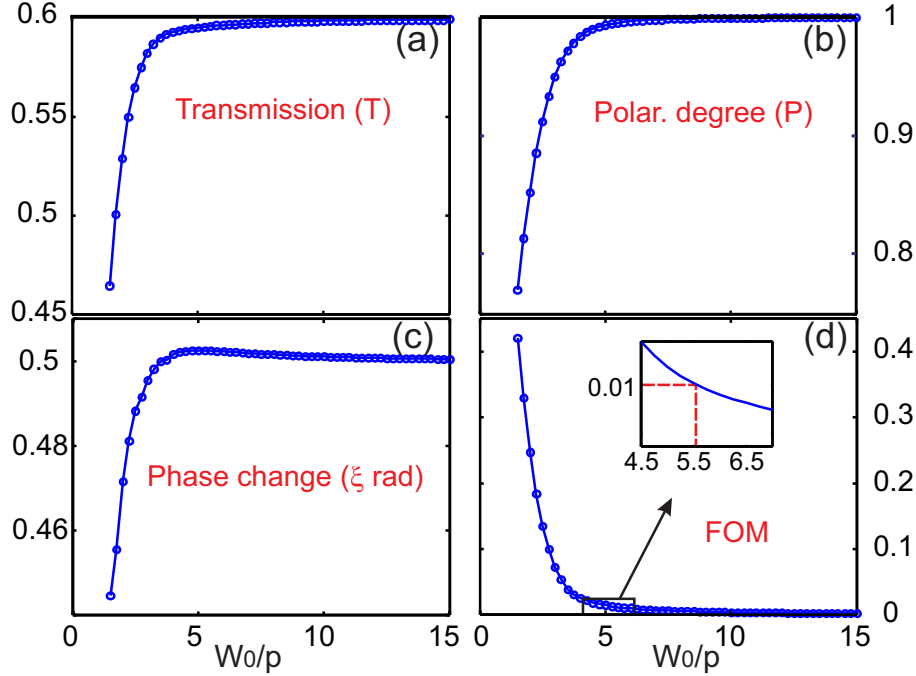}
\caption{Same study as in  {Fig. \ref{demionde}} but for the quart-wave plate operating at $\lambda=1.182p$.}
\label{w0variequart}
\end{figure}

\begin{figure}[ht]
\centering
\includegraphics[width=12cm]{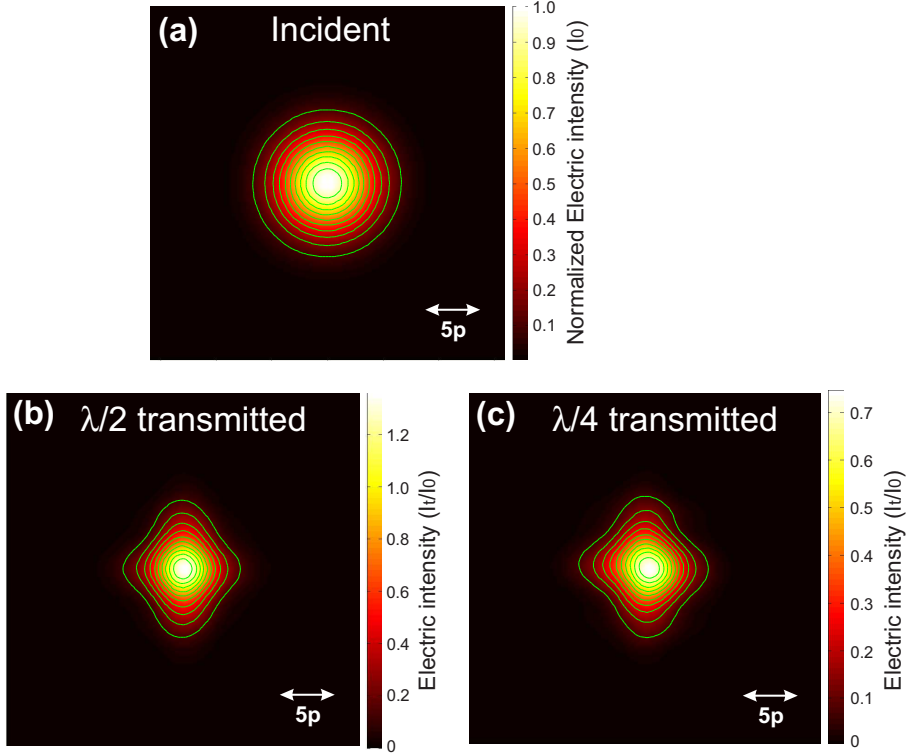}
\caption{Incident (a) and transmitted electric field intensities in a transversal plane located at $Z_0=30p$ from the MAP in the case of a $\lambda/2$ (b) and $\lambda/4$ (c) plates.  The total dimension of each figure is $32p\times 32p$. The waist of the beam is located at $Z_0=0$  { and its value is fixed to $W_0=2p$}.}
\label{gaussiendemi}
\end{figure}

The quarter-wave plate is obtained in ref. \cite{baida:prb11} by only modifying the dimensions of one rectangular aperture. Thus, after optimizing the transmission coefficient, we fix here $a_y=0.658p$ instead of $a_y=0.653p$ in that reference. All the other geometrical parameters are kept the same as for the half-wave plate. The operation wavelength is equal to $ {\lambda_0}=1.182p$. This small modification of the parameter $a_y$ leads to more efficient transmission coefficient of  {$T_{\text{pw}}=0.5995$} instead of $0.57$ and to  {$\xi_{\text{pw}}=0.5001\pi$} and {$P_{\text{pw}}=1$} for a plane wave illumination under normal incidence. Figure \ref{w0variequart} shows the variations of the three parameters $T, \xi$, $P$, and of the  {$FOM$} as a function of the beam-waist $W_0$ value when this MAP ($\lambda/4$ plate) is illuminated by a linearly polarized Gaussian beam. The criterion of {$FOM\leq0.01$} is met for a minimum beam-waist of $W_0=6p$. The condition is quite similar to  {the half-wave plate's one} demonstrating again the high robustness of such  {metasurface} to act as anisotropic plates in the domain of THz or microwave.
To illustrate effect of the angular filtering that can occur for a highly focused Gaussian beam ($W_0=2p=1.67 {\lambda_0}$), we present  {in}  {Fig. \ref{gaussiendemi}(a)} the spatial distribution of the incident electric intensity ($|E_i(x,y,z_0)|^2$) in comparison with the transmitted electric field intensity ($|E_t(x,y,z_0)|^2$) calculated at a distance $z_0=30p$ from the output side of the MAP for half-wave plate [ {Fig. \ref{gaussiendemi}(a)}] and for quarter-wave plate [ {Fig. \ref{gaussiendemi}(b)}]. As it can be clearly seen, for both cases, the transmitted beam is distorted and does not present a 2D Gaussian shape. Nevertheless, for the half-wave plate, the transmitted beam provides more intensity ($1.5$-fold increase) at its maximum than the incident beam even if the total transmitted power ($T=0.81$) is smaller than  {a plane wave one}. This means that the MAP behaves as a lens and focuses the beam at a distance around $Z_0=30d$  {due to the angular filtering operated during the transmission process. Indeed, this transmission depends on the angles $\theta$ and $\psi$ - i.e. for each plane wave - differently from one to another aperture. Both high- or low-pass spatial frequency filters can take place and lead to diverging or converging beams depending on the structure transmission properties. In addition, for a small beam-waist ($W_0$) value, the symmetry of the transmitted beam is broken due to the fact that only few apertures are under illumination.}  To gain {further} insight into evolution of the beam spatial profile, we add two movies corresponding to the evolution of the transmitted electric intensity distribution (on the right of the movies) in a transverse plane as a function of the distance $Z_0$ varying from $100p$ to $p$ for both the $\lambda/2$ (See \textbf{Visualization 1}) and the $\lambda/4$ (See \textbf{Visualization 2}) plates in comparison with the same distribution without the plates (on the left of the movies). As expected, the transmitted field is completely distorted according to the transmission properties that {depend} on the angle of incidence.

In conclusion, we have developed an original tool that allows us to test the robustness of a {metasurface with respect to its proprieties (transmission, reflection, anisotropy) through a  {$FOM$} that can be adapted to its expected properties. This tool is very versatile and can be used to study the effect of any physical parameter such as the angle of incidence, the spatial shape of the beam or the azimuthal angle. Presently, the cases of MAP made in perfect electric conductor behaving as a $\lambda/2$ or $\lambda/4$ plate are studied to point out the robustness of their properties (efficiency and anisotropy) with respect to beam size. Fortunately, we find that $6\times6$ periods $\equiv 5\times5 \lambda^2$ are needed to recover the full properties of a plane-wave illumination. These results are essential for the design of real finite structures and to adapt the experimental conditions so as to achieve optimized results and take maximum advantages of the metasurface properties. The presented methodology is general and can be extended to beams with arbitrary polarizations and angular spectrum distributions.}
\section*{Acknowledgments}
We acknowledge financial support by the Labex ACTION
program (Contract No. ANR-11-LABX-0001-01).
We are grateful to D. Van Labeke for fruitful discussion and helps in the implementation of the monomodal modal method.

\end{document}